\documentclass[12pt,titlepage]{article}
\usepackage{amsmath}
\usepackage{amsfonts}
\usepackage{amssymb}
\usepackage{amsthm,amsmath}
\usepackage{amssymb,latexsym}
\usepackage{amscd,hyperref}
\usepackage[noend]{algpseudocode}
\usepackage{algorithm}
\usepackage{breqn}
\usepackage{diagbox}
\usepackage{enumitem}
\usepackage{cite}
\usepackage{geometry}
\makeatletter
\renewcommand{\fnum@algorithm}{\fname@algorithm}
\makeatother
\usepackage{color}
\usepackage{graphicx}

\title{Combining COVID-19 Vaccination with Social Distancing Measures \thanks{Thanks to Professors Sotiris Vandoros of the King’s Business School, King's College and Nicos Christodoulakis of the Athens University of Economics and Business for several insightful comments and suggestions. Of course errors and omissions are the sole responsibility of the author.}}
\author{Evangelos F. Magirou \\Athens University of Economics and Business\\efm@aueb.gr}
\date{August 16, 2021}
\begin{document}
\maketitle
\begin{abstract}
We analyze an optimal control version of a simple SIR epidemiology model that includes a partially specified vaccination policy and takes into account fatigue from protracted application of social distancing measures. The model assumes demographic (age related) categories, and is otherwise homogeneous. Maximum capacity for vaccination is exogenous and the authorities select its allocation to the categories. They  can also adopt measures to uniformly diminish the contact rate between infected and susceptible individuals at an economic cost.  Total or partial immunity is modeled, while the contact rate is allowed to exhibit seasonality.  We apply optimal control methods to minimize a cost index while guaranteeing that health system capacity is not exceeded.  A reasonable parameter selection leads to policies specifying that vaccination priority should be given to the category with a higher demand for the limited health care resources regardless of cost variations among categories.  Vaccination priority reverses if some demographic category has significantly higher mobility while for some parameter values vaccination alternates among categories. Optimal social distancing policies exhibit seasonality, and hence reducing susceptibles below the average spontaneous disease extinction level does not necessarily lead to a repeal of social distancing.
\end{abstract}
\section{Introduction}
The balance between measures to reduce the spread of a virus and the desire to safeguard social and economic activity is quite delicate, as evidenced by the events since the beginning of the pandemic.  The tradeoffs were  assessed in numerous publications using optimal control methods.  Among the first were \cite{NBERw26981}, where a lock-down intervention is incorporated in a SIR model, including the probabilistic occurrence of a vaccine and \cite{Cowles2229} where properties of the optimal social distancing policy are proven. In \cite{Charpentier2020} the same methods are employed to assess the combination of several intervention modes. Our previous work \cite{magirou2020optimal} was in the middle ground between economics and health policy, stressed the seasonality character of the pandemic and came up with policies that a posteriori proved on the right track as they anticipated the need for timely resumption of social distancing by early autumn 2020 and their unabated continuation throughout the winter months\footnote{Seasonality has been included in several models in the literature because the specific disease is airborne and thus more easily transmitted indoors. As the weather improves during the summer, people become more mobile and are less likely to gather indoors.}.  

The arrival of effective vaccination occurred earlier than was expected in the beginning of 2020, but capacity limitations and vaccine avoidance still pose a serious obstacle to the final relief from the pandemic. The modalities of vaccination and its coordination with social distancing were analyzed using optimal control in \cite{Grundel2020.12.22.20248707} which is based on the authors' earlier compartmental model \cite{grundel2020testing}, in \cite{chhetri2021optimal} where existence of optimal controls was examined, in \cite{libotte2020determination} where a multi objective formulation was presented, while political decision making parameters were considered in \cite{PortugalPandemic}. In \cite{Kontoyiannis} the results of a simulation-based evaluation of several policies for vaccine roll out are reported using a detailed epidemiological model.  We will extend these and other similar work by incorporating seasonality in  disease spreading parameters, taking into account vaccination refusal, modeling fatigue from mandatory social distancing and examining the dependence of optimal policy on health system capacity and category mobility. The population fraction unwilling to be vaccinated is an important parameter, and in health systems with limited intensive care facilities it is unavoidable to continue social distancing even after the reduction of susceptibles below the average level required for spontaneous disease extinction.  Incorporating weariness results in reducing the intensity of measures mainly in the middle of the horizon, but as expected has little effect in low capacity health systems.  

An important finding is that vaccination priority depends mainly on the ICU demanded by each demographic group, to a lesser extent on its transmission characteristics and to an even smaller degree on its economic and disease related costs.  It is of interest to see whether the policies derived in our simple model persist in more complex deterministic or stochastic models.  Needless to say  dynamic parameter estimation and  state identification is of paramount importance for the implementation of optimization policies.  In the same spirit, principles of Model Predictive Control are employed in \cite{MPC} and \cite{WatkinsNP20}.

We present the formulation and solution methods for the model in Section \ref{Model}. Computational results for the Cases examined are in Section \ref{Cases}, and a short Conclusion follows.  Appendix \ref{SecParEst} presents the selection of disease, cost and model parameters while \ref{GradAlgor}  gives an informal description of the algorithm used.

\section{Model Formulation and Solution Methods}\label{Model}
\subsection{Model Formulation}\label{modelform}
We consider the standard  SIR epidemiological model of W. O. Kermack and A. G. McKendrick \cite{Kermack},  \cite{HethcoteSIAM} including  population related, demographic categories and endowed with the potential for controlling the contact parameter, a partially controlled vaccination policy and a modeling of fatigue from non pharmaceutical interventions. Population categories are denoted by the indices $j,k$. Thus let $S_j(t),I_j(t),R_j(t)$ be the number of individuals in a population of size $N(t)$ that are in compartments referred to as \textit{Susceptible, Infectives and Removed} at time $t$ in each population category $j$.  We consider a short horizon, and hence assume a constant category size $N_j(t)=N_j$  the total population being $\sum_j N_j=N$.  We work with the corresponding fractions $s_j(t)=S_j(t)/N$, $v_j(t)=I_j(t)/N,\;n_j=N_j/N$, the removed fraction being determined by  $s,v$ as $n_j-s_j(t)-v_j(t)$.  For convenience we denoted the fraction of infectives by \textbf{$v$} instead of \textbf{$i$}.  

The dynamics of an epidemic  with homogeneous mixing of the population are determined using the following standard assumptions: 
\begin{enumerate}
	\item The rate of newly infected in category $j$ by those in $k$ is $\lambda_{jk}(t)s_j(t)v_k(t)$ 
	\item The rate of removal from the infectives in category $j$ is $\gamma_j v_j(t)$, $\gamma_j$ constants.
	\item The rate of immunity loss of those removed is  $\delta_j (n_j-s_j(t)-v_j(t))$ who then revert to the susceptibles in the same category.
\end{enumerate}

In the above equations $\lambda_{jk}(t)$ is the contact rate, which is the average number of adequate contacts in category $j$ per infective in $k$ and unit time at instance $t$ \cite{HethcoteSIAM}, $\gamma_j$ is the removal (recovery plus death) rate, a constant. The possibility of reinfection i.e. immunity loss at rate $\delta_j$ is included in the model so we are dealing with a controlled SIRS model, although in our calculations we do not consider reinfections setting $\delta$ to zero. Vaccination at a rate $w_j(t)$ is applied to the susceptibles of category $j$ and immediately shifts them to the Removed compartment. Consequently, the  epidemic dynamics are 
\begin{equation}\label{SIR}
\begin{array}{l}
\frac{ds_j}{dt}=-\sum_{k}\lambda_{jk}(t)s_j(t)v_k(t)-w_j(t)+\delta_j(n_j-s_j(t)-v_j(t))\\
\frac{dv_j}{dt}=\sum_{k}\lambda_{jk}(t)s_j(t)v_k(t)-\gamma_j v_j(t).
\end{array}
\end{equation}
More detailed models in the literature  include latent infectives (Exposed) that accelerate the contagion rate (SEIRS). We assume that this effect can be incorporated by the proper choice of contact rates.  We allow rates  to depend on time, modeling thus seasonal variations in virus spread.  The vaccination  is modeled in a rudimentary way as transferring susceptibles to the removed at a rate $w_j$.  In order to model the situation of a population fraction refusing to be vaccinated, we introduce categories of susceptibles $s^R_j(t)$ that are not to be vaccinated and that once infected revert to those accepting vaccination.  These categories follow the equations
\begin{equation*}
\frac{ds^R_j}{dt}=-\sum_{k}\lambda_{jk}(t)s^R_j(t)v_k(t).
\end{equation*}
\noindent It is easily shown that the total infectives follow \eqref{SIR}.  We use the detailed equations in our calculations but for ease of exposition make no reference to those refusing vaccination when examining the optimization model.

We assume that we can select the vaccination intensity in each category subject to an exogenous upper bound on the vaccinations, i.e. $\sum_j w_j(t)\le W_{max}(t)$.  More accurate formulations would include a \textit{vaccinated compartment} that will switch to the removed one at a certain rate, in the meantime acting as susceptibles with better disease response as in \cite{Grundel2020.12.22.20248707}.  A separate category that has received the first dose of the vaccine can be introduced to assess a policy of exclusive first dose vaccination versus a two dose one \cite{Kontoyiannis}.  In the literature, for instance in \cite{Grundel2020.12.22.20248707},  vaccination is expressed as a fraction of the corresponding category, which is convenient since the infected population can never go negative. In the same reference a constraint is placed on total vaccines available at all times, and there is no bound on the vaccination rate. By contrast we insist on always observing the appropriate bound, which in turn causes numerical difficulties.

 We assume that the mitigation, suppression and any other policies mentioned say in \cite{Imperial} can be represented by a single, scalar control variable $u(t)$ with values in $[0,1]$, as in \cite{NBERw26981}. This is an oversimplification: one could study modes of intervention $k=1,..,n$ and consider a vector control $[u_1,..,u_n] $ of the corresponding instrument's intensities \cite{Charpentier2020}.  However all proposed interventions consist of reducing the contact rate so we expect to get some insights from the scalar control case.   We thus model the effect of a level $u$ on the contact rate by the expression
$$ \lambda_{jk}(t)=\lambda^o_{jk}(t)(1-u(t)).$$

The contact rate before a social distancing intervention $\lambda_{jk}(t)$ has a seasonality which (as in \cite{magirou2020optimal}) we represent it by  $\lambda_{jk}(t)=\lambda^o_{jk}(1+k_{seas}sin(2\pi t)),\, \lambda^o_{jk},k_{seas}$ constants.  The intervention effectiveness was  assumed linear multiplicative, and we thus finally express the controlled contact as
\begin{equation}
	\lambda_{jk}(t)=\lambda^o_{jk} (1+k_{Seas}sin(2\pi t) )(1-u(t)).  
	\end{equation}
The inter-category contact coefficients $\lambda^o_{jk}$ determine to a great extent the model's behavior, and by a proper selection one can describe "super-spreader" categories, which is crucial in determining vaccination priority. Data on contacts among categories are extensive in the literature \cite{SocContactPatterns} and we present in  Appendix \ref{contactparms} how to parametrize them in accordance with the assumed contact characteristics.

At every distinct time interval $[t,t+\Delta t]$ we assume a cost proportional to the fraction of infectives.  The cost will consist of the reduced output of those infected that exhibit symptoms, the cost of medical service required (perhaps in addition a discomfort cost) and finally a cost for fatalities.  In \cite{NBERw26981} these costs are in economic terms, and we will follow this approach in our parameter selection.   We thus consider a cost element  $m_j(t)v_j(t)\Delta t$ for each category $j$, the time dependence in the cost coefficient $m_j(t)$ reflecting changes in treatment effectiveness and cost. 

It is important to implement the limited capacity of the health system.  We assume that of those infected $v_j$ in category $j$ a fraction $a_j$ exhibits severe symptoms and requires the use of scarce health care facilities and thus their sum should not exceed an exogenous capacity $V_{max}(t)$,  which can be time dependent to reflect capacity changes.  We must adjust the social distancing parameter $u$ and the vaccination schedule so that the infected $v_j(t)$ always satisfy the constraint $\sum_j a_j v_j(t)\le V_{max}(t)$ and the cost (to be specified next) is minimized.  Alternatively as in \cite{magirou2020optimal}, we  append to the cost integral the penalty term $D\sum_ja_jv_j(t) \exp\left(M(\sum_j a_jv_j(t)-V_{max}(t) )\right)$, where the parameters $M,D$ are to be selected so that the cost is small if the infectives do not stress the system but rises steeply if they do. Each alternative has computational difficulties, so we implemented both, obtaining almost identical results.

The intervention intensity $u$ is assumed to impose a cost which is convex since simple mitigation policies have sub linear costs, while suppression type measures have costs that increase more than proportionally.  Thus we consider a control cost of the form $Au(t)^n \Delta t$, usually quadratic.  

We incorporate a cost term reflecting weariness from lengthy interventions which is important in policy making and thus introduce a fatigue index $z(t)$ which accrues while measures are in effect but decreases to some extent when they are relaxed, exhibiting memory loss.  The fatigue index has the dynamics
\begin{equation}
	\frac{dz}{dt}=b_fu(t)-d_fz(t)
\end{equation}
with parameters $d_f,b_f$.  A penalty term $Bz(t)^m$ is included in the cost  function.   Fatigue increases for measures with intensity in excess of $d_f\frac{z(t)}{b_f}$  and conversely. We expected that this could lead to periods of intensive measures followed by relaxations as practiced by several countries, but we failed to observe such an oscillatory effect, see the computations in Section \ref{Cases}.

We consider the model for a fixed horizon $[0,T_H]$, with an instantaneous cost as described in the previous paragraphs and in addition a terminal cost depending on the final state which is important because just minimizing the period cost would lead to a postponement of infections to the horizon's end.  Assuming that costs are additive in time and are to be discounted at a rate $\rho$ the total  cost is given by the expression:
\begin{small}
\begin{equation}\label{totcost}
	\begin{split}
		\int_0^{T_H}\exp({-\rho \tau}) \left\lbrace \sum_{j}v_j(\tau)m_j(\tau)+ D\sum_j (a_j v_j) \exp (M(\sum_j a_jv_j-V_{max})) +Au^n(\tau)+Bz^m \right\rbrace d\tau \\ +\sum_j m^{term}_j v_j(T_H)\exp(-\rho T_H).
	\end{split}
\end{equation}
\end{small}

 For each category $j$ we denote by  $m_jv_j(t)$ its overall cost per unit time and by $m^{term}_jv_j(T_{H})$ the terminal cost, the parameter assessment to be presented in Appendix \ref{SecParEst}.  We are interested in computing a policy consisting of  intervention and vaccination programs $u(t),w_m(t),w_y(t)$, $t\in [0,T_H]$ to minimize \eqref{totcost}. If policy measures are to be selected to satisfy capacity constraints, the penalty coefficient $D$ will be set to zero.  We only consider open loop policies which however is a step towards feedback, stochastic - adaptive control policies.  
 
 \subsection{Solution Methods}
  Since our model is time dependent we use optimal control algorithms \cite{Bryson} instead of the dynamic programming ones used for instance in \cite{NBERw26981}.  For ease of exposition we present only the penalty function formulation.  We write the Hamiltonian including dual variables $\phi^s_j,\phi^ v_j,\phi^ z$ correspond to states $s_j,v_j,z$ respectively:
\begin{small}
\begin{equation}\label{Hamiltonian}
	\begin{split}
		H=\exp({-\rho t}) \left\lbrace \sum_{j}m_jv_j+ D\left(\sum_j a_j v_j\right) \exp\left(M\left(\sum_j a_jv_j-V_{max}\right)\right) +Au^n+Bz^{m} \right\rbrace\\ + \sum_j \left(\phi^s_j \frac{ds_j}{dt}+\phi^v_j \frac{dv_j}{dt}\right)+\phi^z(b_fu-d_fz)
	\end{split}
\end{equation}
\end{small}
The  dual variable dynamics are
\begin{equation}\label{Costateequ}
	\begin{array}{l}
		\frac{d\phi^s_j}{dt}=-\frac{\partial H}{\partial s_j}= (\phi^s_j-\phi^v_j)(1-u)\sum_{k}\lambda_{jk}(t)v_k+\delta_j \phi^s_j
		\\ \frac{d\phi^v_j}{dt}=-\frac{\partial H}{\partial v_j}=(1-u)\sum_{k}\lambda_{kj}(\phi^v_k-\phi^s_k)s_k(t)+ \phi^s_j\delta_j+ \phi^s_j\gamma_j -\\ \quad\quad
		-exp(-\rho t)(m_j+Da_j(1+\sum_{k}a_k v_k M)exp(M(\sum_{k}a_k v_k-V_{max(t)})))
        \\  \frac{d\phi^z}{dt}=-\frac{\partial H}{\partial  z}=-mBz^{m-1}exp(-\rho t)+d_f\phi^z.
	\end{array}
\end{equation}
The partial derivatives of the Hamiltonian with respect to the controls are
\begin{equation}\label{HamPartials}
	\begin{array}{l}
		\frac{\partial H}{\partial u}= \exp(-\rho t)nAu^{n-1}+\sum_{j}\left[(\phi^s_j-\phi^v_j)\sum_{k} \lambda_{jk}(t)s_j(t)v_k(t)\right]+\phi^z b_z
    	\\ \frac{\partial H}{\partial w_j}=-\phi^s_j
	\end{array}
\end{equation}
 The optimal  policy is determined by choosing the values of $u,\,w_j$ minimizing the Hamiltonian \eqref{Hamiltonian}, which is convex in $u$ so we must have $\frac{\partial H}{\partial u}=0$ for an unconstrained control, and hence the optimal social distancing level $u^*$  is given by truncation in $[0,1]$ of the expression 
\begin{equation}\label{OptContr1}
u^*(t)=\left[ \frac{\exp({\rho t})\sum_{j}(\phi^v_j-\phi^s_j)\sum_{k} \lambda_{jk}(t)s_j(t)v_k(t)-b_z\phi^z}{nA}\right] ^{1/(n-1)}.
\end{equation}
The optimal social distancing expression above shows from a policy perspective why it is important to have a good estimates of the product of susceptibles and infectives. 

Vaccination intensities enter linearly  in the Hamiltonian multiplied by the duals of the corresponding susceptibles.  Thus it is almost always optimal to carry out vaccination  exclusively in the category with larger dual variable and at the maximum intensity - a bang/bang control policy.  If the duals are equal for a non zero interval it might be optimal to vaccinate all categories at nonzero levels.  We did not analyze the possibility of such optimal (singular) policies and they did not arise in the calculations. 

The optimal policy is determined in principle by solving a \textit{two point boundary value problem} consisting of the equations \eqref{SIR} and \eqref{Costateequ} with social distancing specified by \eqref{OptContr1}, and the above bang/bang vaccination policy.  The boundary conditions are: at the initial time $t_o$ we are given the values of the state variables $s_j(t_o)=s_{j,o},\, v_j(t_o)=v_{j,o}$ and at the final time $T_{H}$ the dual variables that must attain the boundary values $\phi^s_j(T_H)=0,\,\phi^z(T_H)=0,\,\phi^v_j(T_H)=m^{term}_j\exp(-\rho T_H)$, the terminal cost derivative.  Such problems are difficult to solve numerically (a comprehensive description is in Ch. 7 of \cite{Bryson}) since the dual variables increase backward in time while the state ones decrease. The possibility of using singular paths presents additional complications. We solved the cases presented in Section \ref{Cases} by a  first order gradient search for optimal control problems (Section 7.4 in  \cite{Bryson}) suitably modified to satisfy the state and control constraints.  An informal description of the algorithm is given in Appendix \ref{GradAlgor}, its details are available on demand.

\section{Computational Results}\label{Cases}
We present several calculations that are relevant to Covid 19 policy making.  An analytic derivation of optimality properties is not straightforward and we restrict ourselves to calculations with parameter sets referred to as Cases, selected to reflect policy questions.  A methodology for parameter assessment (as applied to a particular country, Greece) is presented in  Appendix \ref{SecParEst}.

  We will use \textbf{two population categories}, indexed by $y,m$ that correspond to the age groups $[0,49] ,[50+]$. We will use Population Data for Greece in 2020 as compiled by the Hellenic Statistical Service, included in the disease data in Table \ref{ImperData} in the Appendix. The two categories are of almost the same size, with population shares $n_m=42,77\%$ and $n_y=57.23\%$.  
  
  Category dependent economic and related disease parameters that are the same in all Cases are summarized in Table \ref{Parameters}, while those varying among Cases in Table \ref{CaseParameters}.  The model-related parameters (common in all Cases and categories) are
  
\begin{itemize}
	\item  Control Coefficient $A = 1$.
	\item  Discount rate 3\% yearly.
	\item  Seasonality $k_{seas}=80\%$.
	\item  Vaccination capacity: 70\% of the population per year
	\item  Penalty Exponential Coefficient $M=2\cdot10^5$.	
	\item  Penalty Coefficient $D=0.2$.	
	\end{itemize}

The horizon in all Cases lasts for two years unless otherwise mentioned. Time zero is set near the Vernal Equinox, say April 1st, just before massive vaccination starts. The overall contact parameter $\lambda$ takes the value of $120$ in all cases - see Section \ref*{contactparms} for a justification - except in the last where a milder situation with $\lambda=70$ is examined.

\begin{table}	
	\centering
	\caption{Nominal Parameters}  
	\label{Parameters}   
	\begin{tabular}{|l|c|c|c|c|c|c|c|}
		\hline
		Class&Infected&Infected&Vaccine&ICU&Share&Susceptible&Infectives\\
		&Cost&Cost-High&Accept.\%&Demand \%&\%&Initially \%&Initially \%\\
		\hline
		$m$&16.340&24.871&90&2.70&42.77&38.0&0.4\\ \hline
		$y$&1.390&2.353&70&0.05&57.23&52.0&0.5\\
		\hline 
	\end{tabular}
\end{table}
\begin{table}	
	\centering
	\caption{Case Parameters}  
	\label{CaseParameters}   
	\begin{tabular}{|l|c|c|c|c|c|}
		\hline 
		\hspace*{15pt}Case Number&1&2&3&4&5\\
		Parameter Name&  &  & &  &  \\	\hline
		ICU units per $10^4$&3&10&10&1.5&3\\
		\hline
		Fatigue Factor $B$&0&0&2&0&0\\
		\hline
		$\lambda$&120&120&120&120&70 \\
		\hline 
	\end{tabular}
\end{table}

It is interesting that for reasonable parameter variations vaccination policies are dictated by the requirement to satisfy the health capacity constraint rather than by an effort to reduce costs\footnote{A proper analysis of this vaccination robustness property requires studying the dual variable dynamics \eqref{Costateequ} in order to establish inequalities on the susceptibles' duals.}. In particular, for all Cases it is category $m$ - those over 50 - that has vaccination priority. This changes when ICU demand (column 5 in Table \ref{Parameters}) reverses without changing the other table entries.  Then it is the $y$ category - those under 50 - that is vaccinated first in spite of its very low cost per infection (Column 2 of the same Table).  

In order to examine the relation between vaccination and category contact rates, we use a parameter $\zeta$ which characterizes the excess contribution of category $y$ - those under 50 - to transmission speed, see Section \ref*{contactparms} in the Appendix. We take 20\% as the nominal value of $\zeta$.  Then category $m$ is to be exhaustively vaccinated before starting on the other category. Priority switches  only for large  $\zeta$ values of the order of 100\%, a substantial multiple of its nominal value.   An interesting phenomenon is that for some ranges of $\zeta$'s we observe \textit{priority switching}: One must start with category $m$, and then switch to vaccinating $y$ before exhausting $m$, to which one we return next.  We will present several such switching phenomena in the Cases that follow. 
\subsection*{Case 1}
\begin{figure}[ht]
	\centering
	\includegraphics[width=7.5cm, height=6cm]{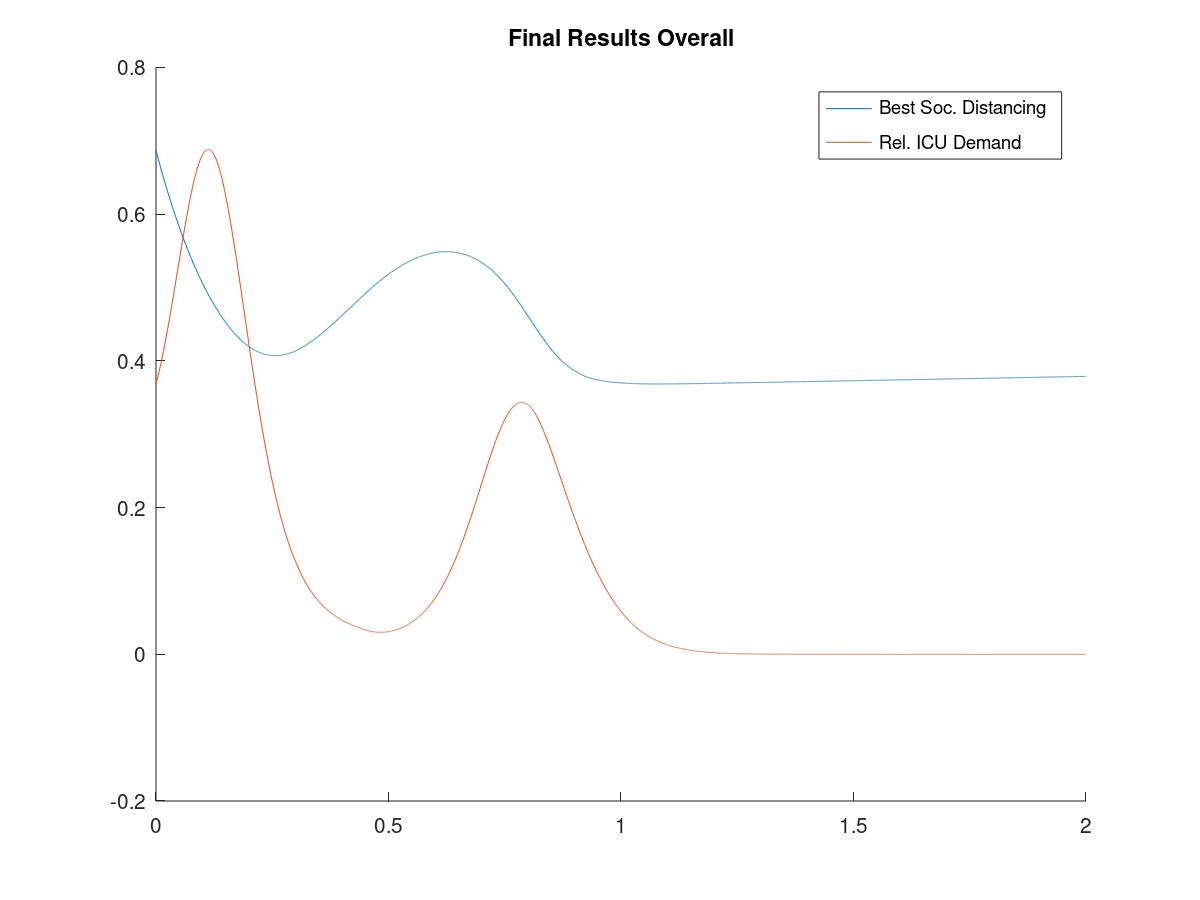}
	\caption{\small{\textbf{Case 1: Intensity of Measures, ICU Demand}} }
	\label{Case_1_overall}
\end{figure}

In \textbf{Case 1} a moderate ICU capacity of 3 units per 10.000 individuals was examined, the remaining parameters being at their nominal values.  Figure \ref{Case_1_overall} shows the "best" social distancing policy (i.e. $u(t)$) and the ICU demanded divided with the ICU capacity (Relative ICU Demand). The social distancing measures follow the seasonal contact variation for a year, starting at a level of about $68\%$, decreasing to $40\%$ in the summer period.  They repeat at a lower level up to about $56\%$ in the winter and then stay at about $35\%$ until the horizon's end, with reduced seasonal variation. Continuation of social distancing was expected since we assumed that about $20\%$ of the population refuses vaccination, which is higher than the level required for spontaneous disease extinction (about $10\%$ in the winter).  We see no obvious explanation for the eventual decrease of seasonal variation. Relative ICU demand peaked at $70\%$ and stayed relatively low throughout the horizon. 
Lengthening the  horizon did not have significant effect on the policies followed, nor changing the cost factors of the infected to the higher values of $m_m=24.871$ and $m_y=2.353$ corresponding to a higher fatality cost ($L=20$).  

\begin{figure}[ht]
	\centering
	\includegraphics[width=7.5cm, height=6cm]{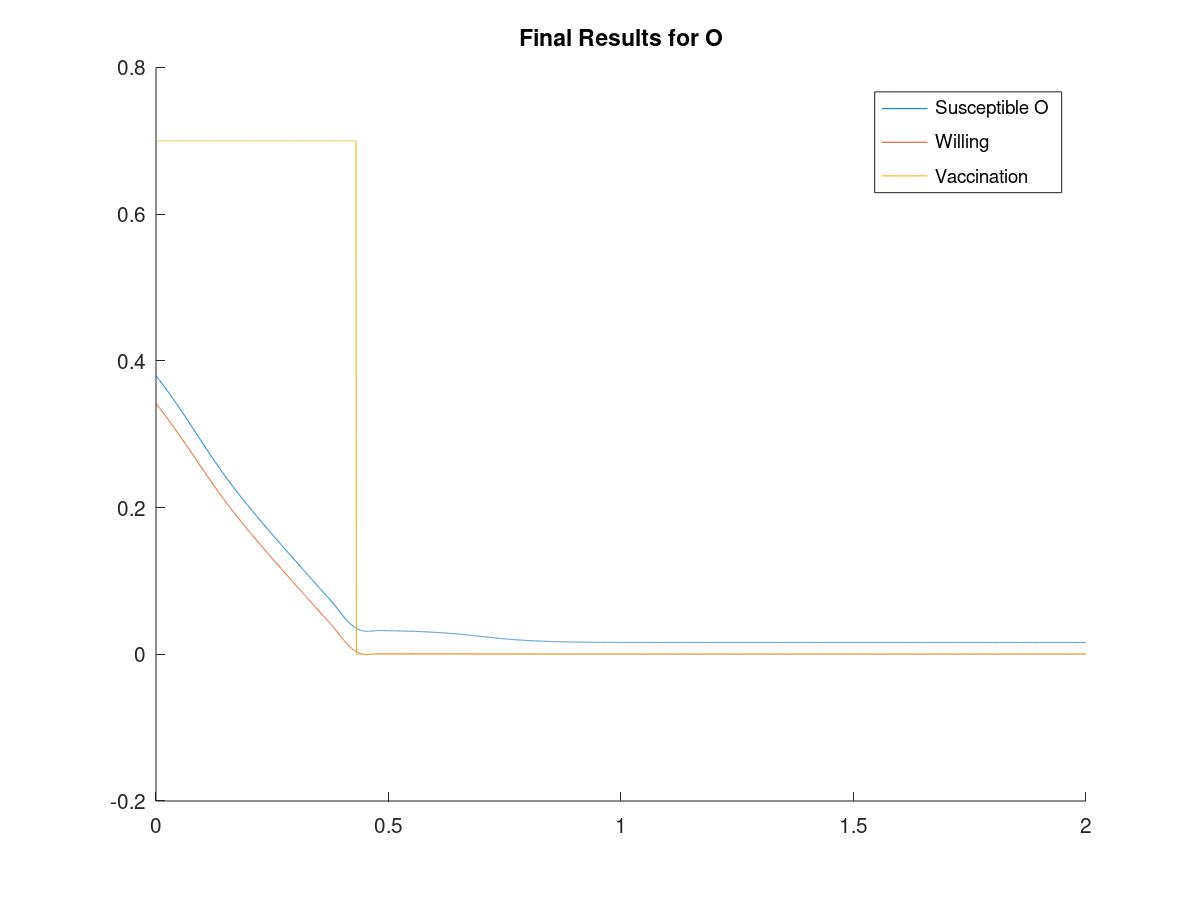}
	\caption{\small{\textbf{Case 1: Vaccination, category $m$}} }
	\label{Case_1_O}
\end{figure}

\begin{figure}[ht]
	\centering
	\includegraphics[width=7.5cm, height=6cm]{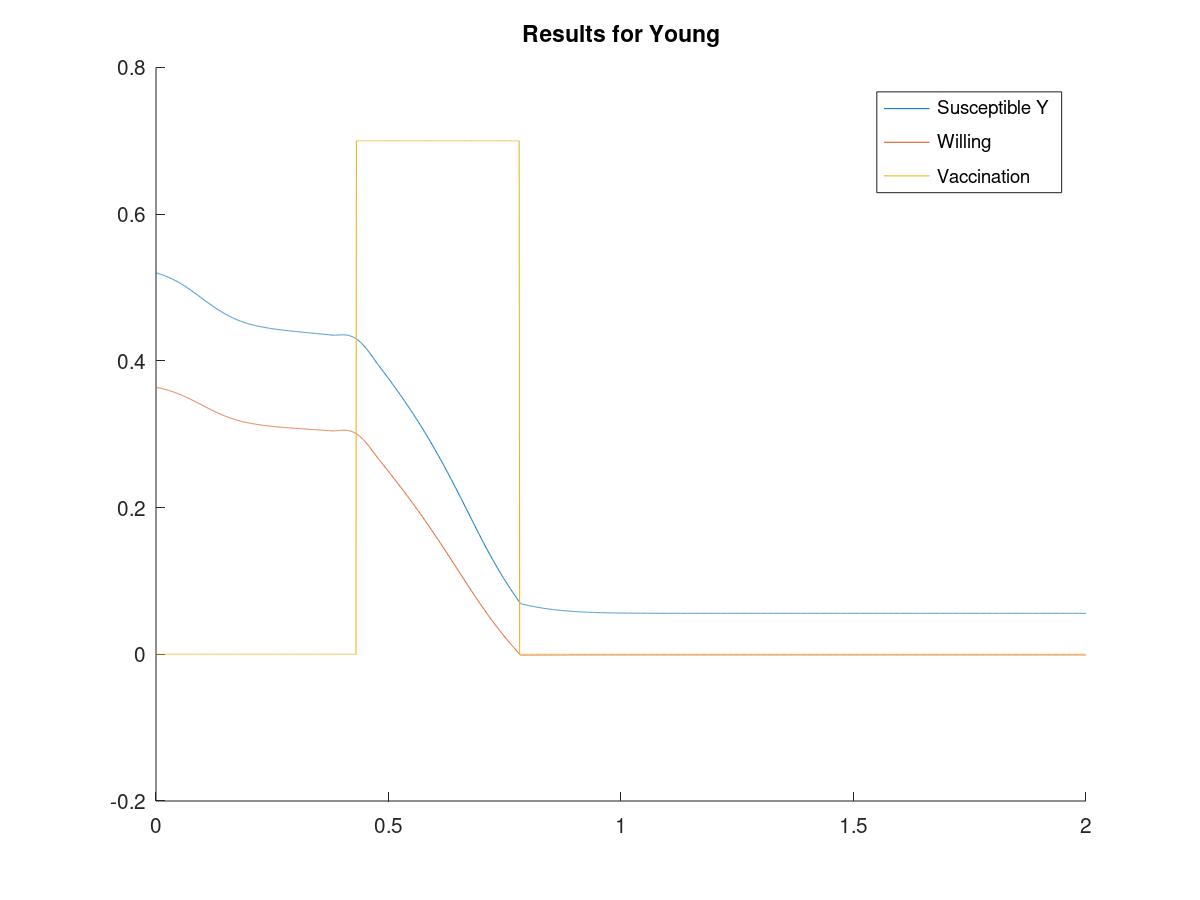}
	\caption{\small{\textbf{Case 1: Vaccination, category $y$}} }
	\label{Case_1_Y}
\end{figure}

Vaccination policies are shown in Figures \ref{Case_1_O} and \ref{Case_1_Y} which show the susceptibles, the susceptibles positive to vaccination and the vaccination intensity in each category.  Vaccination is carried out on category $m$ until the exhaustion of those willing and at the exogenous maximum vaccination rate before starting on category $y$.  Non vaccinated susceptibles, consisting of those refusing vaccination, are considerably reduced but are still sufficient to tax the health system in case measures are lifted. The category contact coefficients are initially at $\zeta=20\%$, and the same vaccination priority persists for value of  $\zeta$ up to about $80\%$.  Then at $\zeta=100\%$ we observe a \textit{switching} policy, starting with $m$, then working on $y$, back to $m$ and then finishing up $y$. When $\zeta$ reaches $200\%$ the policy reverses and category $y$ is vaccinated exhaustively first. In all cases, changing cost coefficients did not affect priorities.

\subsection*{Case 2}
\begin{figure}[ht]
	\centering
	\includegraphics[width=7.5cm, height=6cm]{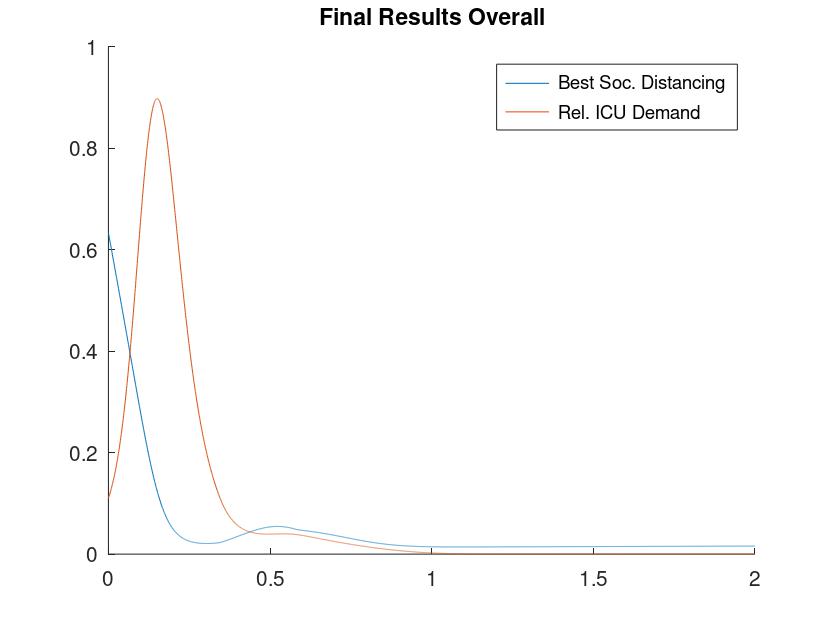}
	\caption{\small{\textbf{Case 2: Intensity of Measures, ICU demand}} }
	\label{Case_2_overall}
\end{figure}

A health system with a high ICU capacity of 10 ICU units for every 10.000 individuals is shown as \textbf{Case 2}, the other parameters as before.  Social distancing  starts almost at the same level as in Case 1, about $63\%$ but decreases drastically in the summer at  $2\%$, then increases mildly to a maximum of $5\%$ next winter and levels at $1.6\%$ for the length of the horizon - see Figure \ref{Case_2_overall}.  Although vaccination is not sufficient for spontaneous disease extinction, very light measures are required to keep the level of infected within the ICU system capacity.

Concerning vaccination policies, again category $m$  is totally vaccinated before proceeding with $y$.  This priority persists even when reversing costs.  As for the dependence on category mobility, the vaccination priority of $m$ is more robust reversing only for $\zeta$ over $1000\%$ with switching among categories for $\zeta$ about $750\%$.  

\subsection*{Case 3}
\begin{figure}[hbt!]
	\centering
	\includegraphics[width=7.5cm, height=6cm]{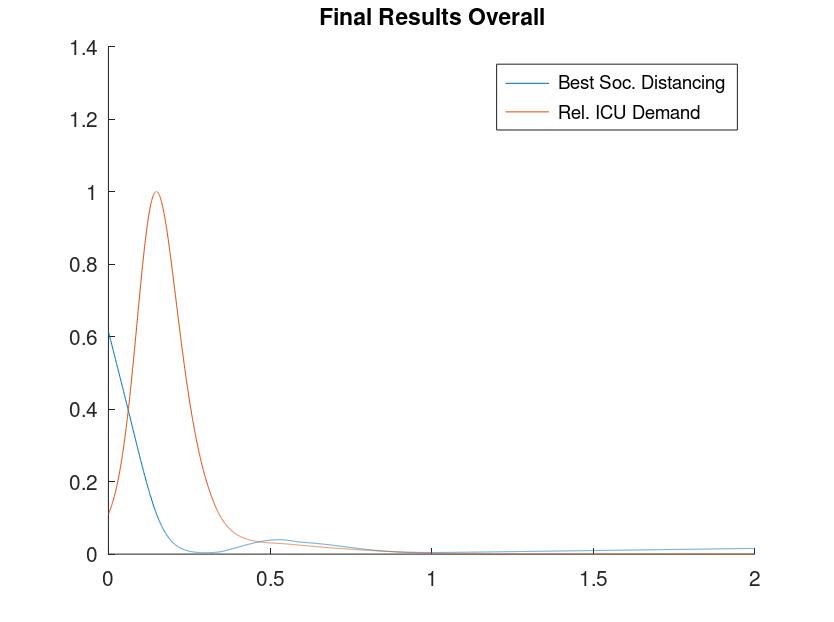}
	\caption{\small{\textbf{Case 3: Intensity of Measures and Infectives}} }
	\label{Case_3_overall}
\end{figure}

\begin{figure}[hbt!]
	\centering
	\includegraphics[width=7.5cm,height=6cm]{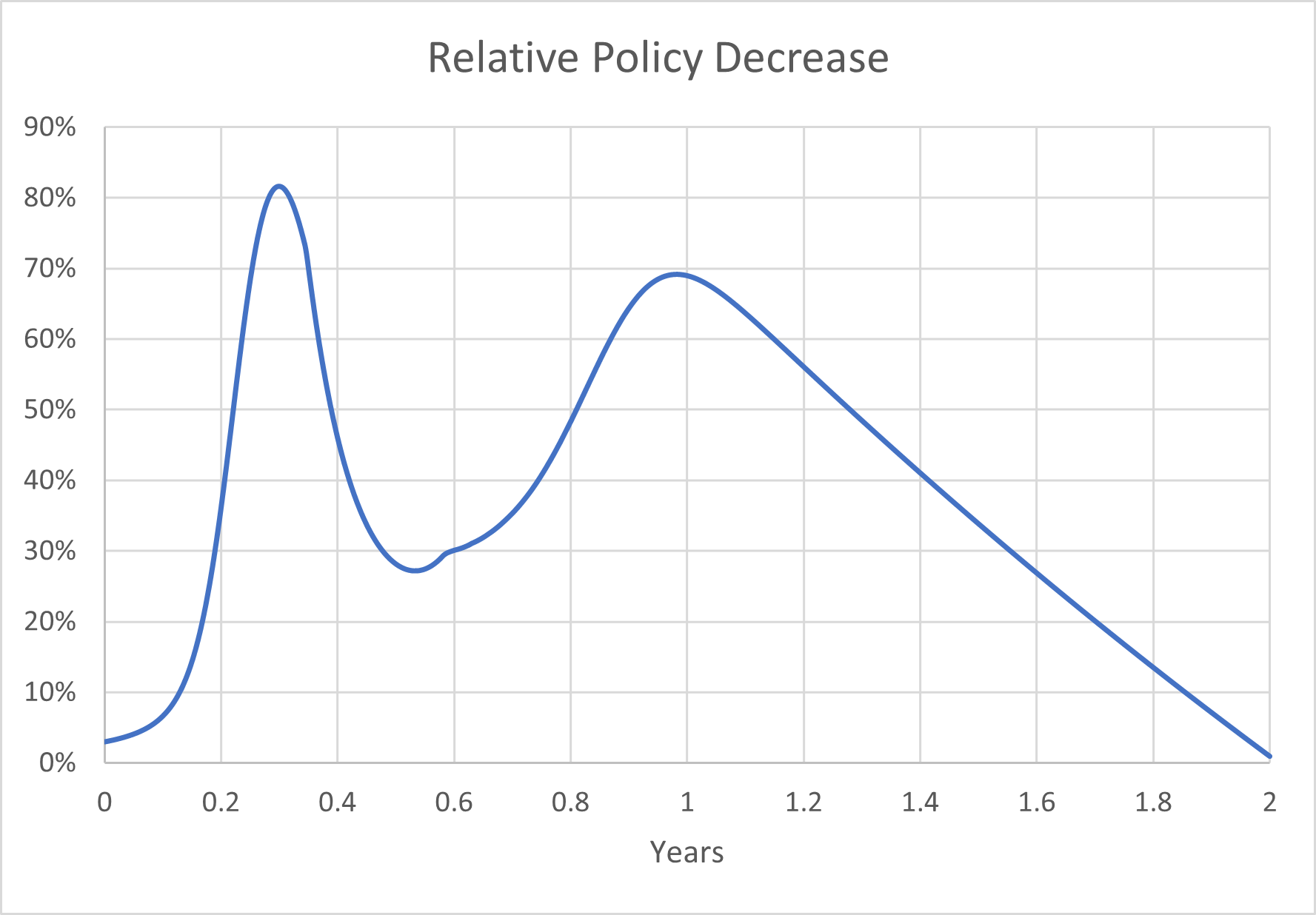}
	\caption{\small{\textbf{Case 3: Relative Decrease in Social Distancing}} }
	\label{Case_3_relative}
\end{figure}

A situation with the high capacity health care system of the previous Case coupled with fatigue that does not subside if measures are relaxed (fatigue parameters $B=2\;b_c=1,\,d_f=0$) is shown as \textbf{Case 3}. The social distancing measures in Figure \ref{Case_3_overall} are slightly lower than those in Case 2, but those infected increase considerably almost exceeding the health system's capacity.  Social distancing decreases are only significant in relative terms ranging from $3\%$ of the non-fatigue $u(t)$ for high social distancing values to $100\%$ in the summer months, see Figure \ref{Case_3_relative}, but they are quite important in view of the considerable ICU demand dependence on social distancing.  Policy does not change appreciably for higher fatigue ($b_f=5$) or when there is memory loss ($B=2\;b_c=1,\,d_f=1$).  Applying the same weariness parameters to the moderate capacity health system of Case 1 had an appreciably smaller effect, decreasing measures by less than $0.2\%$, which is reasonable given the lower ICU capacity.  Vaccination policy has similar characteristics as in the previous Cases.
\subsection*{Case 4}
\begin{figure}[hbt!]
	\centering
	\includegraphics[width=7.5cm, height=6cm]{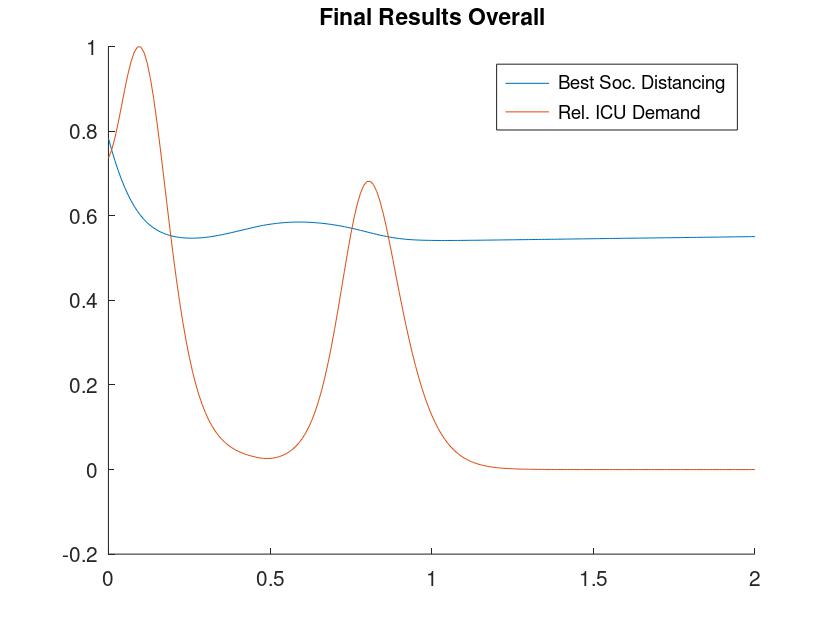}
	\caption{\small{\textbf{Case 4: Very Low Health Care Capacity}} }
	\label{Case_4_overall}
\end{figure}

This case considers a health system of low ICU capacity at $1.5$  units per 10.000 individuals and the results are presented in Figure \ref{Case_4_overall}.  Social distancing measures are more pronounced starting at $78.3\%$, dropping to $54\%$ in the summer, rising next to $58\%$ and remaining close to $55\%$ until the end, showing minimal seasonality. Capacity is almost exceeded initially but subsides soon.

Vaccination policy is more sensitive to category transmission characteristics.   Thus priority is given to $y$ for $\zeta$ as low as $80\%$ with partial vaccination of categories for a $\zeta$ about $50\%$.  This is significantly lower than in the moderate capacity Case 1, where the corresponding values are  at $80\%$ and over $100\%$ respectively, as well as those of the high capacity Case 2 with $\zeta's$ at $750\%$ and greater than $1000\%$.
\subsection*{Case 5}
\begin{figure}[ht]
	\centering
	\includegraphics[width=7.5cm, height=6cm]{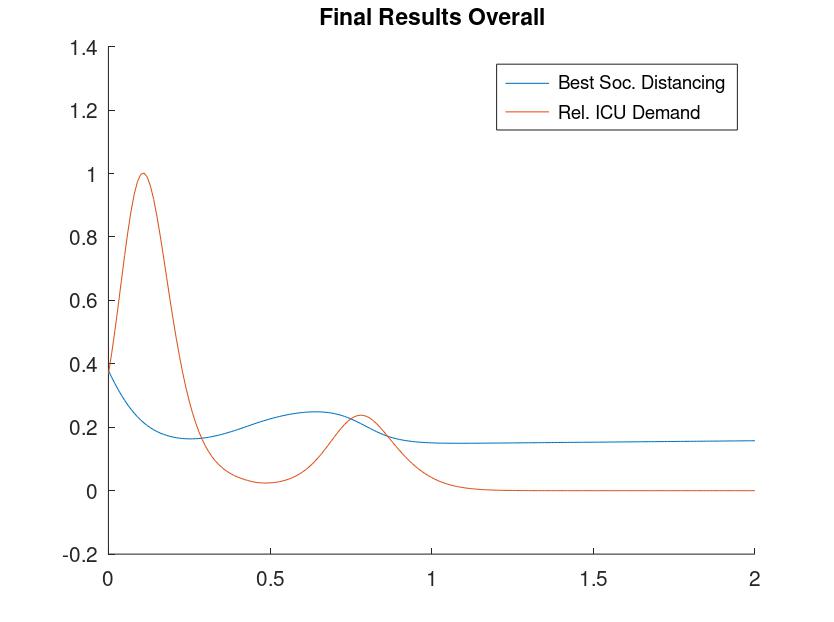}
	\caption{\small{\textbf{Case 5: Intensity of Measures and ICU Demand, Low contact rate}} }
	\label{Case_5_overall}
\end{figure}

In a scenario with a lower contact rate, $\lambda=70$,  \textbf{Case 5}, an average susceptibles level lower than $28.6\%$ seems sufficient for spontaneous extinction, a level that is achieved with the vaccination refusal rates used (10\% for category $m$, 30\% for $y$). However with a seasonality of $80\%$ the susceptibles should be below $28.6/1.8=15.9\%$ at the peak transmission time, and this is not valid with this parameter set.  The results in Figure \ref{Case_5_overall} indeed show that a low health capacity system will impose lower social distancing initially at $38\%$, dropping to $16\%$ in te summer to increase at the level of $25\%$ next winter and then a level of $15\%$ being kept until the horizon's end with minimum seasonal variation. An average contact rate value $L_o\le 55$ guarantees that spontaneous extinction holds for all time after vaccination, and indeed the optimal policy in such a case calls for end of social distancing in a year's time.  
Since the measures are mild, calculations with fatigue parameters did not show significant differences. Thus the relative decrease in social distancing for high fatigue ($B=b_f=2.0$) was less than $4.5\%$ with an average of $2.2\%$.  

\begin{figure}[ht]
	\centering
	\includegraphics[width=7.5cm, height=6cm]{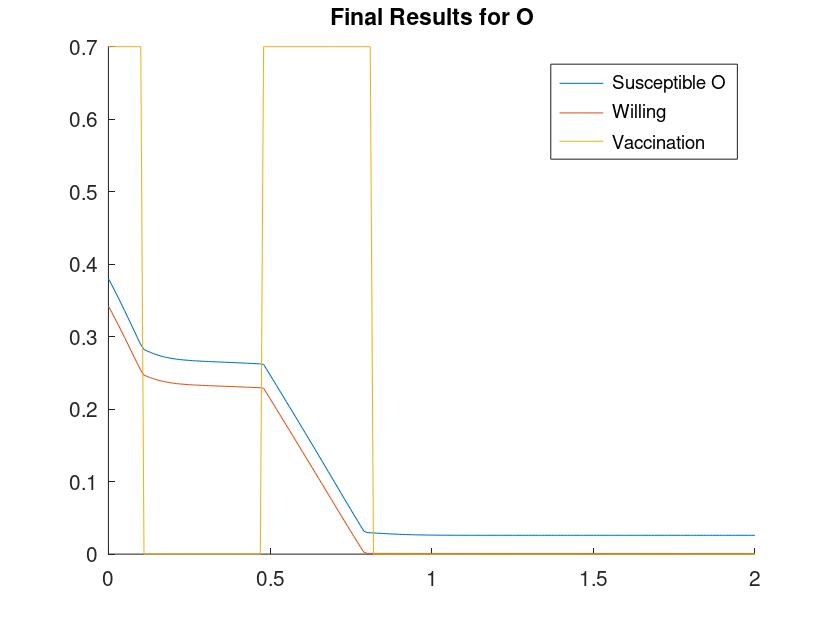}
	\caption{\small{\textbf{Case 5: Category $m$ Vaccination, $\zeta=80\%$}} }
	\label{Case_5ksi80Old}
\end{figure}

\begin{figure}[ht]
	\centering
	\includegraphics[width=7.5cm, height=6cm]{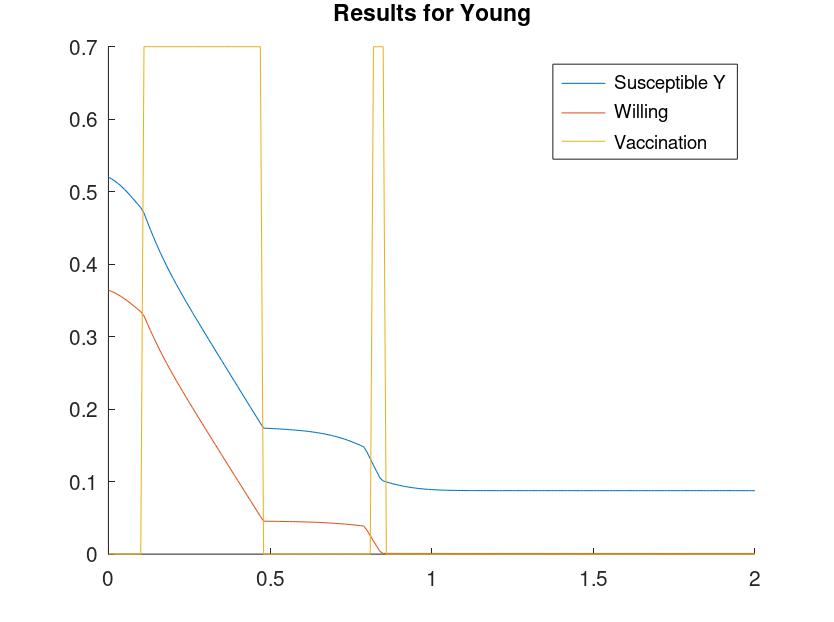}
	\caption{\small{\textbf{Category $y$ Vaccination, $\zeta=80\%$}} }
	\label{Case_5ksi80Yng}
\end{figure}

Category $m$ is vaccinated exhaustively before $y$ for $\zeta$ up $80\%$ when switching occurs, while for $\zeta$ over $100\%$ it is $y$ being exhaustively vaccinated first.  \textbf{Figures \ref{Case_5ksi80Old} and  \ref{Case_5ksi80Yng}} dramatically exhibit the switching effect for $\zeta=80\%$. Thus vaccination switching occurs for the same $\zeta$ as in Case 1, even though the overall contact $\lambda$ parameter there was almost double.  

\section{Conclusions}
 Overall, the findings of this study are quite relevant to policy. While some countries have already vaccinated a large part of their population, most have yet to reach herd immunity, so choosing what groups to focus on to actively promote vaccinations remains relevant. Planning booster jabs, as may be required in the near future, further highlight the importance of such studies. At the same time, vaccination priorities are of great importance for the large number of countries globally that have vaccinated a very small part of their population and need to prioritize based on the small volume of vaccines available. Finally, such modeling applies not only to Covid-19, but also to any future pandemic that requires mass vaccinations. 
 
  It should also be noted that our results are in agreement with the conclusions in \cite{Kontoyiannis} where the optimal strategy is found to be one based on fully vaccinating the elderly/at risk as quickly as possible.  In our computations vaccination priority depends weakly on the infectivity characteristics (contact rates) of the demographic categories. In that sense vaccination decisions take into account a combination of susceptibility and infectivity, in agreement with \cite{Kontoyiannis}, but susceptibility is the dominant factor, while cost differences are not important.  Vaccination policy is "on-off", operating on a single category at every particular time interval.  In most of our computations a category was totally vaccinated before proceeding to the next, but alternating among categories can be optimal although for a limited parameter range.
 
 Our calculations indicate that policies followed by most national health authorities  are sound, although presented from a different perspective.  If one accepts the requirement of staying within the health system's capacity, vaccination priority is necessarily given to those expected to require more resources, while ethical or cost considerations play no role.  The calculations also stress that average herd immunity achievement is not sufficient to stop interventions and that real time state and parameter estimation is of paramount importance.
\clearpage
\appendix
\noindent{\LARGE{ \textbf{Appendices}}} 
 
 \section{Parameter Selection}\label{SecParEst}
 \subsection{General considerations}
 We will select a parameter set consistent with published work and which will serve as a starting point for several scenaria presented in  Section \ref{Cases}.  Given the uncertainty in epidemiological parameters \cite{quanta} one should consider policy recommendations that are parameter insensitive. 
 
 We will use two population categories, while \cite{Grundel2020.12.22.20248707} works with three. Categories will be indexed by $y,m$ and will correspond to the age groups $[0,49] ,[50+]$. We will use Population Data for Greece in 2020 as compiled by the Hellenic Statistical Service, included in the disease data in Table \ref{ImperData}\footnote{https://www.statistics.gr/en/statistics/-/publication/SPO18/-}. The two categories are of almost the same size, with population shares $n_m=42,77\%$ and $n_y=57.23\%$
 
 We will use the disease progress parameters for age groups $[0,9],[10,19],..,[80+]$ appearing in the original Imperial study \cite{Imperial}. Its disease related findings which we use with slight modifications in our parameter selection are summarized below:
 \begin{enumerate}
 	\item One third of those infected will be asymptomatic.  This has be challenged in the literature \cite{Ioannidis}, so erring on the side of caution we will assume fewer asymptomatic cases, namely setting the relevant parameter as $\theta=20\%$. 
 	\item For those infected 5 days will pass until the beginning of symptoms and an average of 18 days are required for recovery. 
 	\item The symptom severity by age group is shown in Table \ref{ImperData}. This provides an estimate of the fraction of infected requiring hospitalization and intensive care.  The recovery period of these categories is larger than average but since they involve a limited population fraction we will not reduce the recovery period of the remaining population
 	\item  Infection Fatality Ratio (IFR) is tabulated by age in Table \ref{ImperData}.
 	\item Those requiring hospitalization need 8 days of hospital stay and 10 days home care regardless of age group. Those exhibiting severe symptoms need a 16 day hospital stay of which 10 days in Intensive Care Units.  Five days pass from the onset of symptoms to hospitalization and a further week until full recovery.
 \end{enumerate}
 \begin{table}	
 	\centering
 	\caption{Disease parameters}  
 	\label{ImperData}   
 	\begin{small}
 		\begin{tabular}{|c|c|c|c|c|c|}
 			\hline
 			Age groups&	Symptomatic cases&Hospitalized cases&Infection Fatality& Population \\	
 			&Requiring &Requiring Critical &Ratio& \%\\
 			& Hospitalization \% & Care \% &\%&\\
 			\hline
 			0-9&0.1&5.0&0.002&9.09\\
 			\hline
 			10-19&0.3&5.0&0.006&10.29\\
 			\hline
 			20-29&1.2&5.0&0.030 &10.49\\
 			\hline
 			30-39&3.2&5.0&0.080&12.44\\
 			\hline
 			40-49&4.9&6.3&0.150&14.92\\
 			\hline
 			50-59&10.2 &12.2&0.600&14.14\\
 			\hline
 			60-69&16.6&27.4&2.200&12.05\\
 			\hline
 			70-79& 24.3&43.2&5.100&9.37\\
 			\hline
 			80+& 27.3&70.9&9.300&7.21\\
 			\hline
 		\end{tabular}
 	\end{small} 
 \end{table} 
 \subsection{Disease and Fatality related costs}
 As in \cite{NBERw26981} we use  costs based on a year's total production $W$ and assume a uniform per capita product $W/N$.  Costs are due to production loss, health care costs and fatalities.   We consider  a small time interval $[t,t+\Delta t]$, compute the sum of costs relating to categories $j$ in it and integrate over the horizon.  We  assume that costs are additive in time and no utility or distribution aspects are of any importance except for discounting. 
 
 Considering first \textit{production loss}, of the $v_j(t)N$ infected in category $j$ some can be asymptomatic or showing slight symptoms that allow them to work (remotely), the rest having to abstain from work.  In \cite{magirou2020optimal} we assumed that $\eta=70\%$ of those in the infected compartment can work (adding asymptomatics to those in an initial or recovery phase), but  here we will use a nominal value of $\eta=50\%$ and thus foregone production is $0.5\sum_{j}v_j(t)W\Delta t$.  
 
 Considering next \textit{non fatality health care costs}, those symptomatic in category $j$ - which as stated earlier are set at $1-\theta=80\%$ -  can be either in home care or hospitalized.   For those in home care we assume a cost equal to $\theta_1=1/2$ of per capita product per unit time and that home care is for the symptomatic period, i.e of 13 out of the total 18 days between infection and  recovery \cite{Imperial}.  We will assign a home care cost for the entire symptomatic period for all infected, even those requiring hospital care by considering hospitalization costs as being over and above those of home care.  Hence the home care cost in that entire category is $(1-\theta)\theta_1v_j(t)\frac{13}{18}W\Delta t=0.289v_j(t)W\Delta t$.
 
 When dealing with hospitalizations we must incorporate age group parameters in the cost coefficients in categories $y,m$. We index age groups by $k$ and categories by $J$ so we can write $k\in J$, and if $N_k$ is the population of age group $k$ and $N_J$ the size of category $J$ we have $\sum_{k}N_k=N_J$, or in terms of the corresponding shares $n_k,n_J$ in the total population $N$, $\sum_{k\in j}n_k=n_J$.  Referring to hospitalizations costs  we consider  them to be again over and above those for home care.  For an age group $k$ the fraction requiring hospitalization among those symptomatic is $h^H_k$ and of those a fraction $h^{ICU}_k$ will require the intensive care facility.  We assume that hospitalization costs are $c_H$ times wage and those for intensive care $c_{ICU}$, again being over and above those in the preceding category.  We will assume that the regular hospitalization length is 8 in a total of 18 days (as in \cite{Imperial}) for all age groups,  while for those requiring intensive care 10 days of ICU and 8 days of hospitalization for an adjusted disease length of 28 days.  Thus  of those infected in age group $k$ the cost due to hospitalization and ICU is $$c_Hh^H_k\frac{8}{18}+c_{ICU}h^H_kh^{ICU}_k\frac{10}{28}$$ multiplied by the fraction exhibiting symptoms. To obtain a coefficient for a category $J$ we assume that those symptomatic in $J,\;v_J(t)$ are allocated in the age groups of $J$ in proportion to their population and thus $v_k(t)=n_kv_J(t)/n_J$.  Then the disease costs in category $J$ in the interval $[t,t+\Delta t]$ are 
 $$ (1-\theta)v_J(t)  \sum_{k \in J}\frac{n_k}{n_J}\left( c_Hh^H_k\frac{8}{18}+c_{ICU}h^H_k h^{ICU}_k\frac{10}{28}\right). $$
 \noindent Using the figures in Table \ref{ImperData} we calculate the summation term as $0.7345$ in category $m$ and only $0.0485$ in $y$.  
 
 \textit{Fatality Cost} parameters are selected in the spirit of  \cite{NBERw26981}.  A uniform loss of life cost is to be added to the production loss for the life expectation of the age group assuming that all years are equally productive irrespective of age.  Such an adjustment would further penalize the aged so we do not implement it. Those exiting the infection stage in age group $k$  in the interval $[t,t+\Delta t]$ are $\gamma_k \Delta t v_k(t)N$, of which  a fraction  $s_k$ succumb. In each group we assume a fixed loss of life cost $L$ and in addition a production loss term $\frac{W}{N}(1-(1+\rho_o)^{-LR_k}))/\rho_o$ which is the present value of yearly production for the life expectation of group $k$, $LR_k$.  Assuming again that the infections in group $k$ are proportional to its population share in category $J$ we have the following expression for the fatality cost in category $J$:
 $$ F_J=\frac{v_J}{n_J}W \Delta t\sum_{k \in J}\gamma_k n_ks_k\left( \frac{L}{\frac{W}{N}}+(1-(1+\rho_o)^{-LR_k}))/\rho_o)\right) $$
 We use the USA actuarial table (Social Security Table: US 2015) and an arbitrary loss of life cost $L/(W/N)=10$ to obtain $F_m=14.816,F_y=0.552$.  To place more emphasis on loss of life, regardless of age, we will also examine doubling the value of $L$ and thus $F_m=23.347,F_y=1.515$.  
 
 Adding all above we will obtain as nominal parameters for the two categories $m_m=0.500+0.289+0.735+14.816=16.340$ and $m_y=0.500+0.289+0.049+0.552=1.390$ , and for the higher cost of life $L/\frac{W}{N}=20$, $m_m=0.500+0.289+0.735+23.347=24.871$ and $m_y=0.500+0.289+0.049+1.515=2.353$ .  Fatality costs are dominant in category $m$ but not in $y$.
 
 \subsection{Demand for Intensive Care Units}
 
 A pressing concern during the pandemic has been the lack of ICU's. A policy goal explicitly articulated has been to impose social distancing sufficient to avoid an ICU deficit.  To implement this constraint in our setting with appropriate coefficients consider age group $k$ at time $t$ of which a fraction $\theta h^H_k h^{ICU}_k$ will require ICU services sometime in the disease's course and specifically for a fraction of $d^{ICU}/d^{Total}$ of its duration.  Assuming as before that the infected portion in age group $k,\;v_k$ will be proportional to the population in its category $J$ we have that the instantaneous demand for ICU for category $J$ is
 $$v_J N \frac{d^{ICU}}{d^{Total}}\sum_{k \in J}\frac{n_k}{n_J}h^H_k h^{ICU}_k=a_Jv_JN$$
 \noindent Using the above data we obtain $a_m,a_y$ equal to $2.70\%$ and $0.05\%$ respectively.  The relevant constraint is then $$a_mv_m(t)+a_yv_y(t)\le ICU/N.$$
 A typical figure for the right hand side bound was about 1 in a population of 10.000 as in Italy at the beginning of the pandemic \cite{Imperial}, but we will consider higher figures in view of pandemic ICU capacity increases.
 
 \subsection{Terminal parameters}
 To assess terminal costs we consider only infections already in effect and not those to occur beyond the horizon.  Assuming an average duration of the disease $\gamma^{-1}_j$, the total non fatality cost of those infected in category $j$ will be  $m_jv_j(t)\gamma^{-1}_j$.  The same expression holds  for the fatality cost and we will thus include a linear cost expression $\sum_j \gamma^{-1}_j m_j v_j(t)$ with coefficients $16.340/20=0.817$ for category $m$ and $1.390/20=0.070$ for  $y$. 
 
 \subsection{Contact Parameters}\label{contactparms}
 A uniform disease duration of 18 days was assumed leading to a  value of $365/18 \approxeq 20$  for $\gamma$ and small values for the reinfection rate $\delta$, since we will consider small horizons.  To simplify the presentation we will not differentiate disease duration among categories, since minor variations do not make appreciable difference in the results.
 
 Based on \cite{Imperial} a value of the overall contact parameter $\lambda=70$ was used in \cite{magirou2020optimal}.  Given the increasing transmissibility of new strains by $50\%-70\%$, as well as the recent Delta variant, a population wide value of 120 will be the new reference point for the contact rate $\lambda$.  Data on contacts between categories is extensive in the literature, for instance \cite{SocContactPatterns}, but here we use contact parameters between categories simply estimated as follows.  First, we assume category $y$ individuals will have $\xi$ more contacts than those of category $m$.  With $\lambda_m,\lambda_y$ the group contacts we have $\lambda=\lambda_m n_m+\lambda_y n_y=\lambda_m(n_m+(1+\xi)n_y)$ and thus $\lambda_m=\lambda/(n_m+(1+\xi)n_y),\; \lambda_y=\lambda(1+\xi)/(n_m+(1+\xi)n_y)$; for $\xi=0.2$ we obtain $\lambda_m=107.7,\;\lambda_y=129.2$. 
 
 We furthermore assume that a member of a category has a probability of contacting a member of the same category which exceeds by $\hat{\gamma}$ that for the other one.  Thus the contacts inside category $m$ are $\lambda_m\frac{n_m(1+\hat{\gamma})}{n_m(1+\hat{\gamma})+n_y}$ and $\lambda_m\frac{n_y}{n_m(1+\hat{\gamma})+n_y}$ with the other class. Taking account conditional probabilities, the infections induced by category $m$ in a unit interval are $$\lambda_{mm}v_m(t)s_m(t)=\frac{\lambda_m(1+\hat{\gamma})v_m(t)s_m(t)}{n_m(1+\hat{\gamma})+n_y}=\frac{\lambda(1+\hat{\gamma})}{(n_m+(1+\xi)n_y)(n_m(1+\hat{\gamma})+n_y)}v_m(t)s_m(t)$$ and $$\lambda_{ym}v_m(t)s_y(t)=\frac{\lambda}{(n_m+(1+\xi)n_y)(n_m(1+\hat{\gamma})+n_y)}v_m(t)s_y(t)$$ to categories $m,y$ respectively.  Similarly for the infections induced by category $y$ we have $$\lambda_{yy}v_y(t)s_y(t)=\frac{\lambda(1+\xi)(1+\hat{\gamma})}{(n_m+(1+\xi)n_y)(n_y(1+\hat{\gamma})+n_m)}v_y(t)s_y(t)$$ and $$\lambda_{my}v_y(t)s_m(t)=\frac{\lambda(1+\xi)}{(n_m+(1+\xi)n_y)(n_y(1+\hat{\gamma})+n_m)}v_y(t)s_m(t).$$assuming the same value for $\xi$ in both categories.   For $\hat{\gamma}=0.2$ we thus obtain $\lambda_{mm}=119.0,\;\lambda_{my}=115.9,\;\lambda_{ym}=99.2,\;\lambda_{yy}=139.1$.
 
 In order to examine the vaccination priority as a function of the transmissibility of each category we assume that the parameters that characterize the excess contacts of category $y$, i.e. $\xi$ and $\hat{\gamma}$ have the common value $\zeta$.   High values of $\zeta$ correspond to higher contribution of $y$ to the transmission.  Its nominal value is assumed at about 20\%.  Our calculations show that in all Cases category $m$ is vaccinated first for $\zeta$  values up to approximately 100\%, but from a point on the other category is to be vaccinated first. 
 
  % We will also examine extreme scenaria where the contact rate of %category $y$ exceeds that of $m$ by $100\%$ and $200\%$, in which case %we obtain  $\lambda_{mm}= %84.369,\;\lambda_{my}=136.966,\;\lambda_{ym}= %70.307,\;\lambda_{yy}=164.359$ in the first case and  %$\lambda_{mm}=61.854,\;\lambda_{my}=150.623,\;\lambda_{ym}=51.545,\;\la%mbda_{yy}=180.748$ in the second.  As shown in the calculations %vaccination priority reverses only in the latter case.
 
 \subsection{Penalty function parameters}
 
 The cost function \eqref{totcost} includes an artificial penalty term to prevent ICU demand to exceed capacity.  This penalty term should be insignificant as long as the ICU demand is below capacity, but high otherwise. The parameters $D,M$ in this  penalty expression $G(D,M)\equiv D\sum_j (a_j v_j) \exp (M(\sum_j a_jv_j-V_{max}))$ are selected according to the following desiderata. First, the derivative with respect to the infected should be small compared to the respective linear cost coefficients when the ICU demand is slightly below capacity.  Second the same derivatives should be very large when there is even a small excess demand.  Thus we consider the derivatives $$ \frac{\partial G}{\partial v_j}=a_jD\exp\left(M(\sum_j a_jv_j-V_{max})\right) \left( 1+M\sum_j a_jv_j\right) $$ first when $\sum_j a_jv_j=V_{max}$ and then when $\sum_j a_jv_j=(1+q)V_{max}$.  We select $D,M$ so that in the first case the $j$ partial derivative is much smaller than the cost coefficient $m_j$ and in the second case they are much larger.  
 
 For convenience we express $D,M$ as $D=\frac{k}{a_m+a_y}$ and $M=K/V_{max}$ (assuming $V_{max}(t)$ is bounded by $V_{max}$).  In fact we only examine cases where the ICU capacity is fixed.  Then the first requirement becomes $kK \ll m_m,m_y$ while the second is $kK\exp(qK)\gg m_m,m_y$.  With the particular values of the cost coefficients for $q=30\%, V_{max}=10^{-4}$ we obtain $k=1/200,\;K=20$ which implies $D=0.2,\;M=2\cdot10^5\;.$
 
 \section{Modified Gradient Optimal Control Algorithm}\label{GradAlgor}

We solve the state constrained version of the optimal control problem stated in Section \ref{modelform} by an ad hoc modification of the standard gradient search algorithm.  Its steps are as follows:
\vspace{.3cm}

\noindent \textbf{Modified Gradient Optimal Control Algorithm}
\begin{enumerate}
	\item Start with arbitrary policies $u^o,w^o_j$ that satisfy control and state constraints. 
	\item \label{Solve}Solve equations \eqref{SIR} for the policy in Step 1 and evaluate its cost.
	\item Given the controls and the state trajectories solve \eqref{Costateequ} for $\phi^s_j,\;\phi^ v_j,\phi^z$  satisfying the appropriate terminal conditions at $T_{H}$ (a differential equation with a single boundary condition) and then evaluate $\frac{\partial H}{\partial u},\;\frac{\partial H}{\partial w_j}$ for $t\in[0,T_H]$
	\item For small $\epsilon$ consider a social distancing policy $u_{new}=u_{old}-\epsilon\frac{\partial H}{\partial u}$ coupled with vaccination to capacity of the category with lower value of $\frac{\partial H}{\partial w_j}$
	\item \label{improvst} Solve \eqref{SIR} using the new controls adjusted to satisfy the constraints on positivity of states and ICU demand.  Evaluate the corresponding cost.
	\item If the cost has not improved repeat Step \ref{improvst}, halving $\epsilon$ until improvement. Note that for small enough $\epsilon$ a lower cost is guaranteed. In principle one should perform a one dimensional minimization along the direction of $\frac{\partial H}{\partial u}$, but this provided only marginal benefits.
	\item Terminate the procedure if the  cost reduction obtained in Step \ref{improvst} is not substantial or the necessary Hamiltonian conditions are approximately satisfied, otherwise repeat Step \ref{Solve}
\end{enumerate}  
A difficulty arises in Step \ref{improvst} since we apply the control adjustments  to the trajectory of the previous iteration, and thus the adjustment will not be valid unless $\epsilon$ is sufficiently small.

\bibliography{BibCovidVacc}
\bibliographystyle{Plain}

\end{document}